# When the Optimal Is Not the Best: Parameter Estimation in Complex Biological Models

Diego Fernández Slezak[1]*, Cecilia Suárez[1], Guillermo A. Cecchi[2], Guillermo Marshall[1], Gustavo Stolovitzky[2]*

1 Laboratorio de Sistemas Complejos, Depto. de Computación, FCEyN, Buenos Aires University (UBA), Buenos Aires, Argentina, 2 Computational Biology Center, T. J. Watson Research Center, IBM, Yorktown Heights, New York, United States of America

## Abstract

**Background:** The vast computational resources that became available during the past decade enabled the development and simulation of increasingly complex mathematical models of cancer growth. These models typically involve many free parameters whose determination is a substantial obstacle to model development. Direct measurement of biochemical parameters in vivo is often difficult and sometimes impracticable, while fitting them under data-poor conditions may result in biologically implausible values.

**Results:** We discuss different methodological approaches to estimate parameters in complex biological models. We make use of the high computational power of the Blue Gene technology to perform an extensive study of the parameter space in a model of avascular tumor growth. We explicitly show that the landscape of the cost function used to optimize the model to the data has a very rugged surface in parameter space. This cost function has many local minima with unrealistic solutions, including the global minimum corresponding to the best fit.

**Conclusions:** The case studied in this paper shows one example in which model parameters that optimally fit the data are not necessarily the best ones from a biological point of view. To avoid force-fitting a model to a dataset, we propose that the best model parameters should be found by choosing, among suboptimal parameters, those that match criteria other than the ones used to fit the model. We also conclude that the model, data and optimization approach form a *new* complex system and point to the need of a theory that addresses this problem more generally.

**Funding:** This work was partially supported by grants from Buenos Aires University (UBA X132/08 and UBACYT M415), Consejo Nacional de Investigaciones Científicas y Técnicas (CONICET) (PIP 5756/06), Agencia Nacional de Promoción Científica y Tecnológica (ANPCyT) (PICTR 184 and PAV 127/03). DFS acknowledges support through a Fundacíon YPF scholarship and the summer internship program provided by the IBM Computational Biology Center. The funders had no role in study design, data collection and analysis, decision to publish, or preparation of the manuscript.

**Competing Interests:** The authors declare no conflict of interest between the contents of this work and the IBM Computational Biology Center, which funded part of this work. The authors confirm that this funding does not alter in any way their adherence to all the PLoS ONE policies on sharing data and materials.

* E-mail: dfslezak@dc.uba.ar (DFS); gustavo@us.ibm.com (GS)

## Introduction

A necessary step in making predictions from mathematical models of biological processes is the estimation of the parameters needed to simulate the model. This is a well studied problem in systems biology, usually addressed utilizing a large variety of approaches [1–5].

Fitting parameters of mechanistic models to experimental data is usually a daunting task [6]. There are several difficulties associated with parameter fitting. One such difficulty stems for the fact that models may display sloppy parameter sensitivities [7,8], whereby some parameters can compensate other parameters, resulting in some arbitrariness in the specification of their values. Another generic difficulty is that different values of the model parameters may be similarly consistent with the data (the problem of identifiability). Yet a third difficulty is that finding the optimal values of the model parameters may require the exploration of a huge space. In this paper we highlight a fourth difficulty usually not discussed when fitting parameters to data. Given a model that is only an *approximate* representation of a system under study, and data extracted from this system, the model parameters that best represent the mechanistic details of the system may not be found by minimizing a cost function. In other words, the parameters at the global minimum of the cost function may not yield the most meaningful parameters from a physiological point of view. In effect, when minimizing a cost function the optimization process can force the search to go to a corner of the parameter space which, while fitting the data exquisitely, yields physiologically meaningless parameters. In view of these difficulties, the question of how to approach the generic problem of searching the parameter space of models that are only approximate representations of complex systems remains a challenging one.

The utilization of mathematical models to describe and predict morphological and physiological aspects of tumor growth [9] has the potential to increase our understanding of tumor development, and holds the promise to suggest new ways to improve the efficacy of therapeutic interventions. Over the last ten years increasingly complex mathematical models of cancerous growth have been developed, in particular on solid tumors, in which growth primarily comes from cellular proliferation [10,11]. The growth of micro-tumors in the avascular stage can be studied using multicellular spheroids as a biological model [12–14]. The multicellular spheroid model [15] is at present considered an excellent *in vitro* model to study complex aspects of tumor physiology, especially those related to therapeutic strategies that cannot be adequately treated by other simpler models [16,17]. This kind of model represents an intermediate level of complexity between *in vitro* monolayer cell cultures and *in vivo* solid tumors. Ward and King [18] proposed a mathematical representation of the processes that describe the growth or remission of an avascular micro-tumor in terms of the nutrient concentration present in the medium, based on a system of nonlinear partial differential equations. The model assumes the existence of a continuum of cells in two possible states: alive or dead. According to the concentration of a generic nutrient, the living cell may reproduce or die, following a saturation kinetics. The division or death of cells implies the expansion or contraction of the tumor volume. The growth of an avascular tumor can be described by the temporal dependence of the radius of its spheroidal volume. This radius is easily accessible by experimentation, and can in principle be used to constrain the parameters of avascular tumor growth models.

We are interested in fitting the parameters of an avascular tumor growth model [18] using time course data. This is a good system to explore the above-mentioned difficulties, associated with parameterizing a model. On the one hand this model has six parameters which is a large, yet manageable number of parameters. On the other hand, the influence of these parameters on the final shape of the growth curve is far from obvious, and therefore there is no simple way to estimate the parameters from exploration of the growth curve. To fit the parameters in this model we did a systematic exploration of the cost function defined as the sum of the squares of the differences between model prediction and time course data, summed over the observed time points. To minimize this cost function we implemented and tested four different algorithms: (1) Levenberg-Marquardt [19,20]; (2) Fletcher-Davidon-Powell [21]; (3) Downhill Simplex [22]; (4) Parallel Tempering [23]. Each of these methods belongs to different families of optimization techniques, described below. We use the massive parallel architecture of a Blue Gene supercomputer to sample the parameter space and characterize the rugged nature of the solution landscape. We found the surprising result that the global minimum of the cost function is not biologically meaningful. This conclusion indicates that the global minimum of the cost function might not be the place to look for the parameters of our model.

In subsequent sections, we will present and discuss strategies to find the parameters of this model, characterize the cost function landscape, and study the parameters they yield in terms of pre-existing biological knowledge.

## Methods

### Model of avascular tumor growth

A detailed description of the mathematical aspects of the avascular tumor growth model to be considered in this paper can be found in the original reference [18], while the numerical implementation of the solution has been described in [24]. In this section, we summarize the basic tenets of the model with the aim of introducing the parameters whose values we want to fit from experimental data. As the experimental model is a spheroid, we can assume a spherically symmetric system whose variables depend on the spatial coordinates only through the radius $r$, i.e., the distance form any point in the spheroid to a fixed center. Three variables determine the model: *i)* the density of living cells, $n(r,t)$; *ii)* the local growth velocity, $v(r,t)$; and *iii)* the nutrient concentration, $c(r,t)$.

After non-dimensionalization and some reordering of the terms [18], the equations for the density of living cells is:

$$\frac{\partial n}{\partial t} + v\frac{\partial n}{\partial r} = \{[k_m(c) - k_d(c)](1-n) - \delta k_d(c)n\}n,$$

where $k_m(c)$ and $k_d(c)$ are the rate of mitosis and death respectively, and $\delta$ is the fraction of the original cell volume that a cell occupies after it dies, i.e., $\delta = V_D/V_L$ with $V_L$ and $V_D$ denoting the volume of a living and dead cell respectively. The dependencies of $k_m$ and $k_d$ on the concentration are given by saturating functions assumed to have the form

$$k_m(c) = \frac{c}{c + c_c}, \qquad k_d(c) = 1 - \sigma\frac{c}{c + c_d}.$$

The parameter $c_c$ represents the crossover concentration above which the rate of mitosis reaches its normalized saturation value of 1. In like manner, the critical concentration $c_d$ denotes the crossover concentration of nutrient below which the normalized death rate saturates to 1 and above which the normalized death rate saturates to $1 - \sigma$; here, $\sigma$ denotes the basal cell death rate parameter, which is independent of nutrient conditions.

The equation for the normalized local velocity is [18]

$$\frac{1}{r^2}\frac{\partial(r^2 v)}{\partial r} = [k_m(c) - (1-\delta)k_d(c)]n.$$

Finally, we will use the equation for the nutrient concentration inside the spheroid. This equation results from a quasi-steady approximation of the reaction diffusion equation ruling the nutrient concentration:

$$\frac{1}{r^2}\frac{\partial}{\partial r}\left(r^2\frac{\partial c}{\partial r}\right) = \beta k_m(c)n,$$

where $\beta$ represents the amount of nutrient consumed on cell mitosis. It is assumed that the nutrient consumption by non-mitotic processes is much smaller than that the nutrients consumed during mitosis.

The initial conditions

$$r_{bd}(0) = 1, \qquad n(r,0) = 1, \qquad c(r,0) = c_{bd},$$

specify that the radius $r_{bd}(t)$ at the boundary of the tumor at time $t=0$ is our unit of lengths and that the initial density of cells $n(r,0)$ is one, that is we start with a unique cell submerged in a medium with nutrient concentration given by $c_{bd}$, the latter being also the nutrient concentration at the tumor boundary.

The boundary conditions are:

$$\frac{dr_{bd}}{dt} = v(r_{bd}(t),t), \qquad c(r_{bd}(t),t) = c_{bd},$$

$$v(0,t) = 0, \qquad \frac{\partial c(0,t)}{\partial r} = 0.$$

The first boundary condition in the first line implies that the boundary of the spheroid moves with the local velocity at the boundary. The second boundary condition establishes that the spheroid is immersed in a medium with nutrient concentration $c_{bd}$. The third and fourth boundary conditions (second line) are respectively that the center of the tumor is not moving (i.e., our center of coordinates is at the center of the tumor), and that the radial gradient of concentration at the center has to be zero by the spherical symmetry.

In summary, this model has 6 free parameters, represented by a 6-dimensional array $\theta = (\delta, \sigma, \beta, c_c, c_d, c_{bd})$. The original model as formulated in [18] had four additional parameters, which we are setting to the values suggested in [18]. By their normalization, our six parameters are contained in the interval [0,1]. Different combinations of parameters lead to one of three possible evolutions: linearly increasing, saturating, and decreasing (Figure S1 in File S1).

### Experimental Data

Multicellular spheroid techniques have been widely used and studied since the 1980's [25–28]. One of the most important measures obtained from these spheroids is the growth curve, i.e., the radius or volume of the spheroids as a function of time. With the experimental technology currently available, this curve can be measured very precisely. In [13,14] the authors measured the growth curve of multicellular tumor spheroids used as paradigms of prevascular and microregional tumor growth and compared the experimental growth curves to many different empirical models of multicellular spheroid growth. The authors concluded that the standard Gompertz curve [29], defined as $r_{bd}(t) = a \exp[b \exp(ct)]$ could not be distinguished from the actual growth curves within the margin of error of the experimental data. Taking into account these observations we used a noisy Gompertz curve to constrain the parameters of our tumor growth model. We generated synthetic growth data from a Gompertz model that was previously shown [13,14] to follow very closely experimental volume growth data with values of $a = 7.6$, $b = -12$ and $c = -0.121$. We also added 5% of noise to the volume growth curve, a level of noise similar to what is observed in real experiments. Figure 1 shows the volumetric growth curve generated using the Gompertz model with 5% of noise (left panel), as well as the corresponding non-dimensionalized radius growth curve (right panel). The latter curve will be used in the remaining of this paper to constrain the parameters of a tumor growth model.

### Parameter Estimation

The parameters included in the tumor growth model are represented by the 6-dimensional array $\theta$. In order to obtain numerical values for our model parameters, we will attempt to minimize the differences between our tumor growth model and the data by finding the minima of the quadratic cost function [30]:

$$\chi^2(\theta) = \sum_{k=0}^{N} (y_k(\theta) - \text{data}_k)^2,$$

where $y_k(\theta)$ is the model's prediction for observation $k$ which depends on the parameters $\theta$, and $\text{data}_k$ represents the experimentally measured data values (in our case, synthetic data) at observation $k$. In our particular application, synthetic experimental observations are taken at times $t_k$, and the value $\text{data}_k$ corresponds to the radius of the spheroid at time $t_k$. The model prediction for the spheroid radius at time $t_k$ is denoted by $y_k(\theta)$, and the sum is over all the time points $t_k$. For future reference, the residuals vector **f** is defined as $f_k = y_k(\theta) - \text{data}_k$.

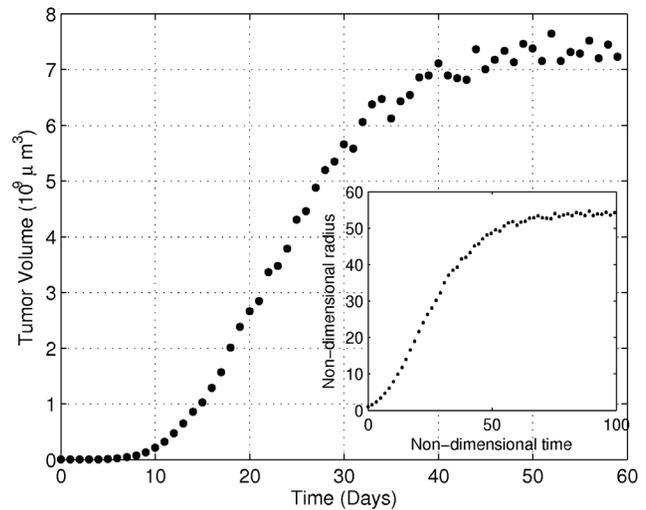

**Figure 1. Data generated using a Gompertz model that fitted previously reported experimental data.** Gompertzian fit to the measured volumetric data with a 5% of noise added. The in-box axis corresponds to the same data, where the non-dimensionalized radius as a function of non-dimensionalized time is illustrated, used as the data in the remaining of this paper.
doi:10.1371/journal.pone.0013283.g001

The methods used for the minimization of the cost function are Levenberg-Marquardt [19], Parallel tempering [23], MIGRAD [31] and downhill simplex [22] (see supplemental section 2 for method details).

Because of the complexity of our partial-differential-equation based model, the simulation of each spheroid growth may take several minutes in a single Blue Gene [32,33] node, the exact time depending on the parameter values. To have a dense sampling of the parameter space, independent runs were implemented and executed in a massively parallel environment. Communication between nodes was implemented using MPI [34]. The parallelization consisted of running the different methods from many starting points simultaneously and parallelizing the independent model evaluations, i.e. Jacobian calculation in LM or MIGRAD and different temperature simulations in PT. To sample the parameter space with the goal of identifying the minima of the cost function, hundreds of thousands of runs were executed. These runs took several months even leveraging the thousands of compute nodes available in Blue Gene. The availability of these resources resulted in a relatively dense evaluation of the parameter space, and the identification of what seems to be the global minimum of the cost function.

### Results

As explained in the previous sections, the model has 6 free parameters $\theta = (\delta, \sigma, \beta, c_c, c_d, c_{bd})$, each bounded in the [0,1] interval. Sometimes we will refer to the parameters as Parameter 1, Parameter 2, etc., with the number indicating the order in the parameter array. We will first show a comparative evaluation of the advantages and disadvantages of the different optimization approaches. For this we will use each method to minimize the cost function $\chi^2(\theta)$ defined in the previous section, where: **(a)** $\text{data}_k$ corresponds to the non-dimensional tumor radius at time $t_k$, given by non-dimensional synthetic data (right panel of Figure 1) and **(b)** $y_k(\theta)$ is the tumor radius at time $t_k$, $r_{bd}(t_k)$ that results from running the avascular tumor growth model with parameters $\theta$.

The sum of squares were performed over the evolution of the radius of the tumor over 100 time points $t_k$ as shown in Figure 1.

Our objective is to determine which parameter sets $\theta$ minimize $\chi^2(\theta)$. A six-dimensional grid of two inner points for each parameter was generated for the initial parameter values from where to start the optimizations. For each of the four optimization methods to be explored, these 64 mesh points were executed in parallel. For the PT method, a first run for sampling the parameter space was executed with 64 different temperatures. After this initial sampling, a second run with very low temperature (all nodes the same temperature) was executed in order to reach the local minima for each parallel replica. The starting points were the parameters corresponding to the minimum cost function value attained in the previous runs.

Figure 2 shows the histograms of the minima of the cost function attained by each of the methods investigated. We observe that many of the minima found by the Simplex Method are orders of magnitude larger than those found by the other methods, with the smallest minimum of the cost function being in the 100's (in the non-dimensionalized units of length squared). In some runs MIGRAD execution had to be interrupted before finishing because it reached the stipulated maximum number of iterations; in these cases, we called "minima" the smallest values calculated up to that point. MIGRAD and PT were able to find similarly good values, with the smallest minima of the cost function being around 20. The LM method clearly found the best minima of the cost function, with smallest values being around 2 (Figure S2 in File S1).

To further study the differences between the LM and PT/MIGRAD minima, we applied Principal Component Analysis [35] (PCA) to the LM solutions. The direction that explains most of the variance has mainly components on the 4th and 6th parameters. Applying PCA to the solutions reached by the others methods with good parameter space sampling (Migrad, PT) yielded two main principal components for which, as was the case for LM, parameters 4th and 6th have the largest components. Thus we plot the projection of the minima sampled by the different methods on the plane given by the 6th ($c_{bd}$) and the 4th ($c_c$) parameters. This two-dimensional glimpse of the six-dimensional parameter space is shown in Figure 3 for all the methods considered.

The Simplex method (top right subfigure) shows a very poor parameter space coverage and a rather high range of cost function values at the attained minima as evidenced by the scale of the color bars to the right of the figure. MIGRAD and PT show a similar tendency toward finding minima in the high range of the 6th parameter values. LM shows the best set of minima with a very significant lower value for the 6th parameter, in contrast with the tendencies in the MIGRAD and PT runs. It is also interesting to highlight the very different search strategies of the different methods. By construction, PT implements a stochastic search strategy, and therefore reaches corners of the parameter space not sampled by the other methods. LM has a good coverage of the parameter space, and it takes a more ordered but yet rather non-local search of the phase space. MIGRAD performs a localized search, more local than PT and LM, but less local than Simplex. In the latter, the original starting mesh of the simulations can be clearly identified, and not too far reaching excursions from the original mesh are explored.

The location and number of local minima shown in Figure 3 suggest that the cost function landscape is a rather rugged one, plagued with local minima. In Figure 4 we explicitly constructed the surface of the cost function as a function of parameters 4 and 6, using all the runs available from all the optimization methods used, yielding a total of more than 100,000 evaluations. This figure shows an extremely rugged landscape with many peaks and valleys permeating the parameter space. This ruggedness is smoothed out by the fact that the value of the cost function shown in Figure 4 is the average over of the cost function values with the same values of parameters $c_{bd}$ and $c_c$, but different values for the other four parameters. We also tried different strategies to represent the cost function landscape, such as plotting the minimum over of the cost function with the same values of parameters $c_{bd}$ and $c_c$. This alternative representation of the cost function yielded a similarly rugged landscape (data not shown).

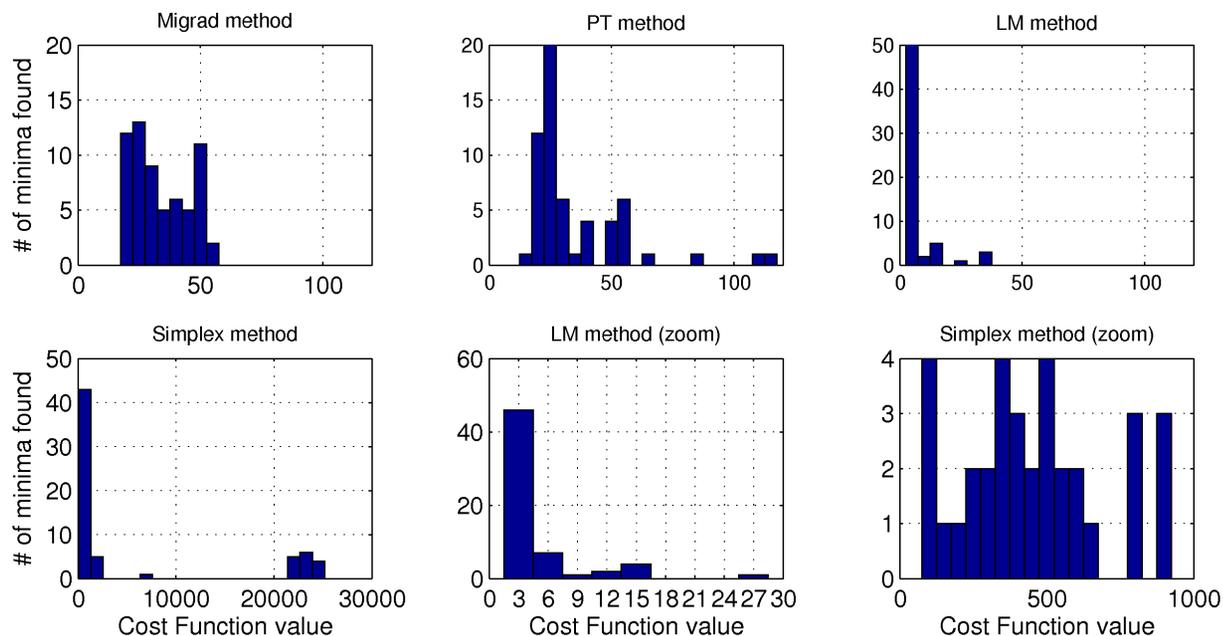

**Figure 2. Histograms of the minima of the cost function obtained using different optimization methods.** In all cases, runs started from a fixed 6-dimensional mesh of 64 points, in which each parameter was evaluated at 2 values. The optimization methods were launched and run until convergence, and the minimum obtained in each run was recorded to create these histograms.
doi:10.1371/journal.pone.0013283.g002

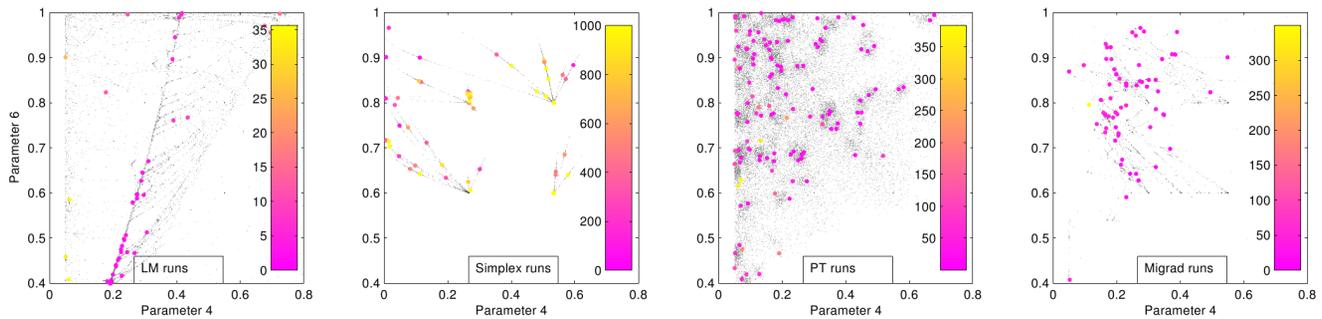

**Figure 3. 6th ($c_{bd}$) vs. 4th ($c_c$) parameters runs for all methods.** Grey points represent individual runs. Colored points are the best cost function values obtained in for each of the 64 initial conditions. The value of the attained minimum for each run is color-coded in the color scale at the right of each subfigure. Note that the color bars have different scales.
doi:10.1371/journal.pone.0013283.g003

Can the ruggedness of the cost function landscape be due to the noisy nature of the data being fitted? It could be conjectured that noiseless data would yield a cost function landscape that is less plagued with local minima. If this were the case, it would make sense to smooth out the data before fitting a model to it. To address this question we recalculated the landscape removing the 5% noise that had been added to the Gompertz curve, and therefore fitting the model to a smooth curve. The results (data not shown) show a landscape with a similar ruggedness as that of Figure 4, indicating that the rough surface cannot be attributed to the noise in fitted data.

To explore the behavior of the parameter space over the other parameters, we show the sampled parameter space projected onto the 5th ($c_d$) and the 1st ($\delta$) parameters in Figure 5 for LM, PT and MIGRAD, the three methods with good minima sampling. Minima found with LM resulted in 5th parameter values very close to zero. Even though MIGRAD and PT show a similar tendency, the minima values reached by these methods are not as low as those obtained by LM. One way to explain this difference is that PT bases its searches on random moves, and the probability of getting in a trough as narrow as $10^{-8}$ (the value found by LM for parameter 5) is very unlikely. On the other hand, as we said before, MIGRAD was interrupted before finishing because the number of iterations performed reached the maximum stipulated. As the MIGRAD method resembles LM, it might be expected that a sufficiently long MIGRAD run would eventually reach similar minima values as LM, albeit in many more iterations.

Even though many of the 64 minima obtained in each method had similar values of the cost function (Figure S2 in File S1), the corresponding parameter values were not necessarily close. In order to understand the structure of the parameter sets found using the different methods, we ordered the parameter sets using standard hierarchical clustering techniques (Figure S3 in File S1). As a metric for the hierarchical clustering we used the correlation between the parameter vectors (normalized inner product), and used average linkage as the method for hierarchical agglomeration of clusters.

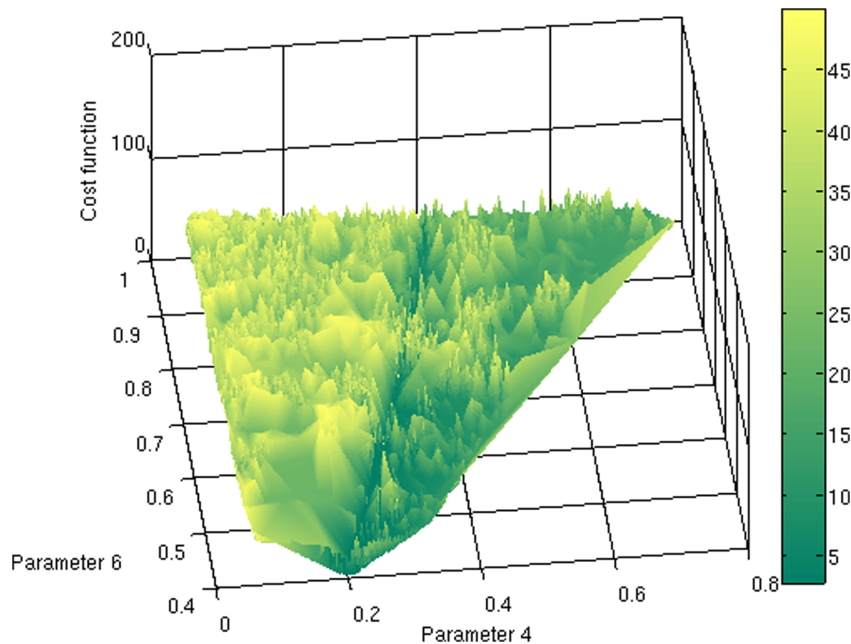

**Figure 4. The landscape of the cost function as a function of the 6th ($c_{bd}$) and 4th ($c_c$) parameters.** The cost function value has been averaged over the values that correspond to the same values of $c_{bd}$ and $c_c$, but for which the other coordinates differed. This data was taken from all runs available of all methods.
doi:10.1371/journal.pone.0013283.g004

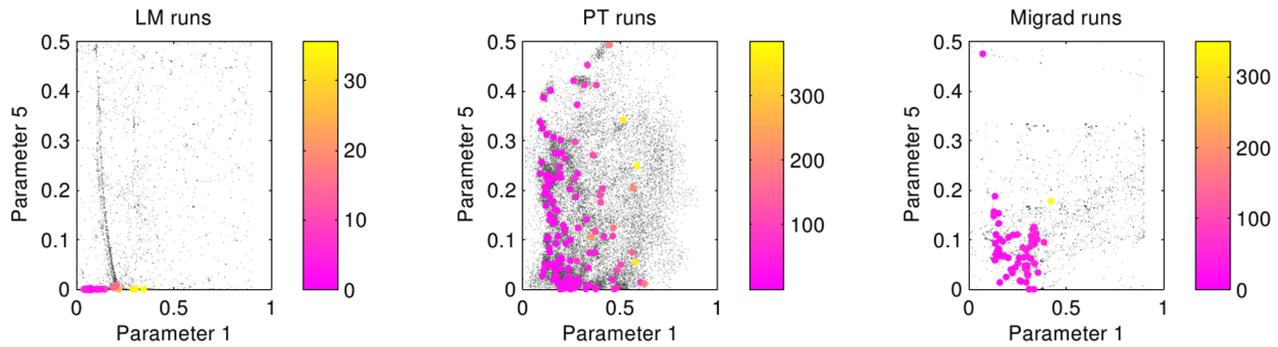

**Figure 5. All optimization runs as seen from the cut of the parameter space through the 5th ($c_d$) and 1st ($\delta$) parameters, for the LM, PT and MIGRAD methods.** Gray points represent individual runs. Color points are the best cost function value obtained in each compute node. The color bars coding the value of the cost function at the minima are in different scales for the different methods.
doi:10.1371/journal.pone.0013283.g005

We divided the sets of minima into seven clusters for each method. The clusters resulting from the LM optimization are tighter and more homogeneous within clusters than those resulting from PT. For each of the seven clusters and each method the mean value of each parameter was chosen as the representative solution (Table S1 and Table S2 in File S1).

Conventional wisdom would indicate that the best parameter set is the one that minimizes the cost function, i.e. the best fit to the experimental data. In the present case, the parameter set that yielded the minimum from LM method would have been chosen. Notwithstanding, growth curves from all minima closely followed the experimental growth curve (see Figure S2 in File S1). Even though we have 100 points to constrain the six parameters of the model, there is still the possibility that the optimization has "forced" the parameters of model to data. We claim that optimality of the cost function is not the only criterion for choosing the best parameters: the best parameters have to be interpretable, and should compare well with their experimentally measured counterparts. The current literature provides very good experimental measurements for $\beta$ and $c_c$ (the 3rd and 4th parameters, respectively). Therefore we checked the consistency of the parameters obtained from our optimization of the cost function with the experimentally measured values of $\beta$ and $c_c$. In Figure 6 we plot the minima obtained by the different optimization methods (solid points) and the cluster centroids (x symbols). In the figure blue and green correspond to LM and PT methods respectively. The biological realistic values reported in literature are shown in the gray box.

Only one minimum centroid for each method was within the biologically feasible values. Surprisingly, neither of these two parameter sets corresponded with the cluster that contained the minimum for either PT or LM. This is a counterintuitive result, as one would expect that the global minimum is the one that optimizes all aspects of the solution. However, the results shown in Figure 6 clearly show that only a few of the minima are biologically realistic, and the global minimum is not amongst these. We shall discuss this issue further in the Discussion section.

Both parameters sets that yield values consistent with the known biology are very similar, except for the 5th parameter which in LM is seven orders of magnitude smaller than in the PT minima. In order to shed some light into this difference we performed sloppiness analysis [7,8] for the model around the parameter sets given by the two centroids. In sloppiness analysis, a slightly modified cost function is used, in which the values $\text{data}_k$ are replaced by the values of the model at the optimal parameter $y_k(\theta^*)$. The Hessian of this modified cost function is computed, and its eigenvalues and eigenvectors are studied. The spectra of eigenvalues for our system had a very wide range of values, with the ratio of the maximum to the minimum eigenvalues being separated by several order of magnitude. The eigendirections corresponding to the large eigenvalues are stiff directions, whereas the sloppy directions in parameter space are those directions in which an excursion doesn't change the modified cost function considerable. Generally, the directions determined by the eigenvectors in sloppiness analysis are linear combination of parameters. In both methods (LM and PT) the minimum eigenvalue corresponded to an eigenvector which was essentially the vector (0,0,0,0,1,0). Therefore, the difference of several orders of magnitude difference in the 5th parameter may be explained by the fact that this parameter corresponds to the most sloppy direction in parameter space, at least in the neighborhood of the centroids.

## Discussion

### When the optimal is not the best

The avascular tumor model used in this work has 6 parameters, each of which represents a physiological mechanism. We fitted these six parameters to data derived from a three-parameter Gompertz curve with added error. Is there any point in trying to fit data that can be fitted with three parameters to a much more complicated model with six parameters? We believe that this parameter-counting argument (six parameters in a model to fit a three parameter Gompertz curve) can be misleading. The avascular tumor growth presented here is a nonlinear partial differential equation in which the solution $r_{bd}(t)$ depends subtly on the equation parameters. If the full model were perfect (to within the error), the optimization should find physiologically reasonable parameters that reproduce the data in an indistinguishable way from the Gompertz curve, regardless of the fact that the data can also be fitted with an empirical model of 3 parameters plus noise.

One of the main conclusions of this paper is that *the notion that model parameters have to be obtained by global minimization of a cost function may be too strong a generalization*. Parameter fitting requires not just brute-force computation but also some strategic thinking. The problem is not so much fitting the data at hand, but rather the ability of the model to make predictions under conditions different from the ones used in the fitting. This is a similar problem as the one faced in machine learning when we want to generalize a classifier to previously unseen data after training it in a training set: a perfect fit to a training set may be due to overfitting, and may result in poor generalization to previously unseen data. In terms of mechanistic models, a perfect fit may result from the fact that the data to be fitted is in the realm of possible outcomes of the model, even with unrealistic parameters. In this section we will elaborate further on these ideas.

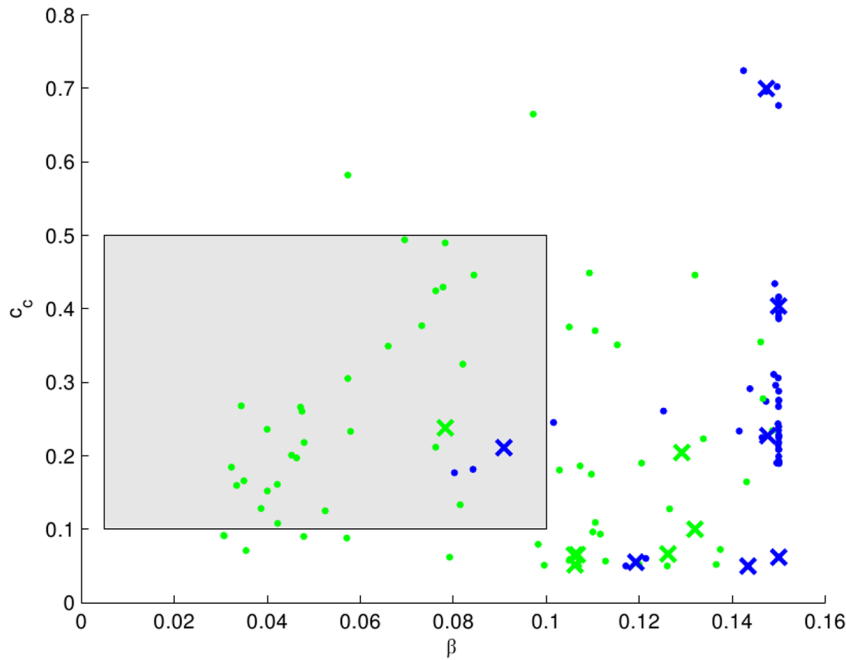

**Figure 6.** $c_c$ **vs** $\beta$ **points of all minima found (dots) and cluster centroids (crosses).** Blue and green represents the minima obtained using the LM and PT methods respectively. The gray box shows the biologically feasible parameter space.
doi:10.1371/journal.pone.0013283.g006

In the problem studied in this paper, we formulated the model of avascular tumor growth based on a set of parameters with a relatively clear interpretation. At least some of these parameters could in principle be contrasted to experimentally measured parameters. In this framework, we have presented a methodological approach for parameter estimation and evaluation of an in-silico model of avascular tumor growth. We evaluated several algorithms to find best-fit parameters, and encountered a proliferation of local minima embedded in a very rough cost function surface. We clearly found a best performing optimization method in LM, which efficiently sampled the parameter space and found what appears to be the global minimum of our cost function.

The optimal parameters obtained by our extensive search of the parameter space was not, however, the best minimum in the sense of making the model interpretable. Indeed, the global minimum consisted of values for the parameters $c_c$ and $\beta$ that were outside of the region where experiments place these parameters. Therefore, even if the fit to the growth curve of the model with the global minimum parameters optimized the cost function, any additional measurement that depends crucially on the value of these parameters would make the model fail to match the data. In other words, in our case, the optimal minimum of the cost function was not the best solution of our problem as it yielded an unfeasible parameter set. The two sets of (cluster centroid) parameters which were consistent with the literature showed a reasonable fit, albeit not optimal. Indeed, the relative error of the model at those parameter values fall outside the error bars of the experimental data (Figure S4 in File S1).

This seemingly paradoxical situation in which the optimal parameter set may not be the best solution to our problem can be illustrated with a simple example. Suppose that we have a set of measurements that depend on a non-dimensionalized time $t$ according to $\text{data}(t) = t + t^3$. Suppose that our model is of the form $y(t) = \alpha t + \beta t^2$. It is clear that our model, while qualitatively correct, is not exact. Continuing with our example, assume that we have collected data from time $t = 0$ to time $t = T$ (where $T$ is also non-dimensional). The cost function that we need to minimize is:

$$\chi^2(\alpha, \beta; T) = \int_0^T dt [y(t) - \text{data}(t)]^2$$
$$= \int_0^T dt [(\alpha - 1)t + \beta t^2 - t^3]^2.$$

The global minimum of this function is obtained at parameter values $\alpha^*$ and $\beta^*$ given by the relations:

$$\alpha^* = 1 - \frac{2}{5}T^2, \qquad \beta^* = \frac{4}{3}T.$$

The best value for the parameter $\alpha$ should be 1, as this parameter represents the importance of the linear coefficient of the data whose linear coefficient is 1. Indeed $\alpha^*$ is close to 1 for small values of $T$, which is where the model best represents the data. However, if we sampled for a longer time, the paradoxical situation results that our estimate of the parameters worsens. It is clear that the parameter $\beta$ while trying to capture the curvature of the data, is parameterizing the wrong dependence (a $t^2$ dependence rather than the $t^3$ dependence of the data in our example). For a long sampling time $T$, the parameter $\beta$ grows linearly with time, as it tries to compensate the smaller curvature of the model in order to fit the data. The parameter α also has to compensate in order for the model to match the data at the high range of $t$. So, if $T=0.5$, the values of $\alpha^*$ and $\beta^*$ are 0.9 and 0.67, very close to the actual values of 1 and 1 respectively. However, if $T=5$, we would have $\alpha^*=-9$ and $\beta^*=6.67$: clearly the linear term is assuming negative values, very far from the reality of the actual data. The optimization process has forced the parameters to fit the data and in so doing, the parameters lost their interpretability of being the linear coefficient and a coefficient related to curvature.

Our intention in presenting this analytical example is to show that when the data is not in the realm of the results that the model can produce, the optimal of the cost function may not yield a meaningful set of parameters. Our simple example makes this point obvious, as we know the functional form that represents the data.

There is another lesson that we can extract from this simple example. It is clear that the optimal parameter values depend on the range of the data to be fitted. Therefore, one simple test of the sanity of a model's optimal parameters is to fit the data at different time points (or in data sets with different range of values). If the optimal parameters dramatically depend on the range of the data to be fitted (such as the dependence of $\alpha^*$ and $\beta^*$ on $T$), then something is wrong with the model.

In their interesting discussion on modeling in systems biology, Cedersund and Roll [36] suggest that a model can be viewed under three epistemological lenses: 1) A model is used as an instrument (as a means) to obtain a certain prediction (instrumentalism). Typically this is the approach that is used when data is modeled using generic statistical methods such as regression; 2) A model is to be the "perfect" representation of the real system (direct realism), as is the quest when theoretical physicists seek the ultimate laws of nature. This means that the "perfect" model will not only be able to give accurate predictions of the measurable system output, but also will provide an accurate description of all the components and processes involved in the generation of this output. 3) An intermediate view exists between 1) and 2) according to which a model yielding good predictions on a diverse data set could be expected to contain some degree of correlation between its mechanisms and the corresponding mechanisms of the real system (critical realism). In this view, a model is considered as a simplification of the true system that only captures some of its aspects, and one therefore has to be careful when drawing conclusions about what these aspects might be. For the avascular tumor model discussed in this paper, some of the simplifications were the assumed perfect spherical symmetry, the Michaelis-Menten growing behavior, the disregard for the discrete nature of the cellular composition of the tumor and the mechanical stresses that cells exert on each other, etc. Cedersund and Roll suggest, and we agree, that of these three approaches (i.e. instrumentalism, direct realism and critical realism), it is the last option that best describes modeling for systems biology. When our models are simplified version of reality and therefore not exact, the best parameters (in the sense of being physiologically meaningful) may not be found by optimization.

If parameter estimation cannot simply rely on cost function optimization, how are we to choose our parameters to determine our models? We believe that the optimization of a cost function, even a cost function with regularization, is only one side of the coin in the fitting process, and that experimental design [37] has to be considered simultaneously with the optimization process. If we have independent data sets probing different regimes of a system, a good strategy may be to take all the local minima solutions that fit the first data set to within an approximation (but not just the global minimum, assuming it can be found), try each of those solutions on the new dataset, and choose the parameter set that fits the best to the second data set. Some experiments will determine some parameters better than others, so a reasonable strategy for parameter fitting is to produce independent experiments that constrain different parameters. Lumping all the experiments in a single cost function may not be the best approach to find parameters from data, as parameters values may be strained until the data at hand is fitted. It may be preferable to fit the model with a subset of experimental data, and contrast the resulting minima with the rest of the experimental data, specially reserved for this purpose. This approach is akin to the cross-validation technique used in statistical learning.

## A formalization

To further explore how the approach sketched in the previous paragraph can be applied, we next describe a formal methodology to choose model parameters in the avascular tumor growth model or any other biological system. We start by sampling the cost function $\chi^2(\theta, D_1)$ with a first constraining data set $D_1$. In the case of the avascular tumor growth model $D_1$ was the spheroid radius-versus-time data. Rather than just choosing the parameter $\theta_1^*$ that minimizes the cost function as the "true" parameter values, we postulate that the "best" parameter values are contained in the set $S_1(\varepsilon)$ of parameters that render the cost function not larger than $\varepsilon^2$. The choice of $\varepsilon$ will depend of the experimental error $\sigma$ in the data, and on the error $\varepsilon_M$ in the fit due to simplifications of the model, plausibly following the relation $\varepsilon^2 = \sigma^2 + \varepsilon_M^2$. $S_1(\varepsilon)$ is thus defined as

$$S_1(\varepsilon) = \{\theta | \chi^2(\theta_1^*, D_1) \leq \chi^2(\theta, D_1) \leq \varepsilon^2\}$$

It may be expected that a simplified model at its best parameters (i.e., realistic and amenable to predict the results of new experiments) can represent the data, albeit with a relatively large $\varepsilon_M$. If the model is based on first principles, one could expect $\varepsilon_M$ to be very small, as the theory should account for the data very faithfully. A principled value of $\varepsilon_M$ (and thus of $\varepsilon$) is hard to determine, but an empirical way for its computation will be discussed below.

The next step in the process is to choose a subset of the parameters contained in $S_1(\varepsilon)$ that are consistent with other data sets. In the case of our avascular tumor growth model, possible additional data sets can be extracted from the following experiments: 1) changing the nutrient concentration of the surrounding medium of the spheroid to obtain different saturation levels of growth curves and 2) measuring the necrotic core size, for example by histological or immunohistochemical markers. These experiments will generate additional data sets $D_2, D_3, \ldots$, etc. In general, the nature and design of these additional experiments will depend on the system being studied. The $n$-th experimental data set $D_n$ will determine the $n$-th plausible parameter set $S_n(\varepsilon)$ defined as

$$S_n(\varepsilon) = \{\theta / \chi^2(\theta_n^*, D_n) \leq \chi^2(\theta, D_n) \leq \varepsilon^2\}$$

where $\theta_n^*$ is the parameter that minimizes $\chi^2(\theta, D_n)$. If we want the $\varepsilon$ to be the same for all plausible parameter sets, the cost functions have to be normalized by the number of experimental data points. If we have a total of $k$ datasets, we want to find those parameters that satisfy the constraints imposed by all datasets. This is simply the intersection of the plausible parameter sets imposed by each data set:

$$S(\varepsilon) = \bigcap_{i=1}^{k} S_i(\varepsilon)$$

We choose $\varepsilon$ to be the minimum such that $S(\varepsilon)$ is non-empty.

If $S(\varepsilon)$ has only one element, we take that to be the best parameter. If it has more than one parameter, we take the best parameter value as the one that minimizes the cost function $\chi^2(\theta, D_1, D_2, \ldots)$ constructed using all the available data sets. If the value of the resulting $\varepsilon$ is too large, then the fits of the model through the data sets are very poor, indicating that the model is too rough to represent the data, and needs to be refined. If the model can produce an acceptable $\varepsilon$, with reasonable fits through the data, then the model is a good representation of the actual system in the realm of the mechanisms probed by the experiments that produced the datasets $D_1, D_2, \ldots, D_k$.

It is interesting to make a parallel between cost function optimization and thermodynamics. In a thermodynamic system at zero temperature, we expect the system to be found in a microstate that minimizes the internal energy. If we increase the temperature, the system will be able to attain other energy states, and at some of these energies there may be a large number of compatible microstates. At a given finite temperature, the system will not be typically found at the microstate that attains the minimum energy, but at a set of microstates that negotiate a balance between having as small an internal energy as possible, and as many microstates available as possible (i.e., as large an entropy as possible). The internal energy at a finite temperature is no longer the global minimum of the internal energy. Rather, the internal energy adjusts itself to be at a value in which the system minimizes its free energy. The parallel with our parameter estimation problem is as follows. The space of model parameters is equivalent to the set of microstates in a thermodynamics system. When $\varepsilon$ is zero (zero temperature), i.e., there is no experimental noise, and the model is a perfect representation of reality, we expect the actual parameters of the model to be those that minimize of the cost function (as the microstate of the thermodynamics system at zero temperature corresponds to the minimum of the internal energy). However, when $\varepsilon$ is non-zero, many parameter values are compatible with the same value of the cost function, and we cannot any longer claim that the right parameter set is the one that minimizes the cost function (as we cannot say that at finite temperature the right microstate of the thermodynamic system is the one minimizes the internal energy). In the case of the parameter estimation problem, it may be possible to define a modified cost function that is the equivalent of the free energy in the thermodynamic system, and which is minimized at the set of parameters compatible with a given $\varepsilon$. This discussion, however, is beyond the scope of the present paper.

### Final conclusions

We conclude that, in the absence of a model based on first principles, parameter estimation cannot just rely on cost function optimization. We have seen that it is necessary to double-check the parameters to identify possible runaway solutions given the complexity of the solution space. Our independent assessment of the parameters with data not used for cost function optimization allowed us to further restrict the minima and reach a compromise between cost function optimality and biological plausibility. We submit that this approach should be considered as an important part of the process of assigning parameters to complex biological models.

### Supporting Information

**File S1**  Supplemental figures with the corresponding text.
Found at:   doi:10.1371/journal.pone.0013283.s001   (0.70  MB PDF)

### Acknowledgments

We are grateful to Jeremy Rice, Lior Horesh and Nikhil Bansal for fruitful discussions.

### Author Contributions

Conceived and designed the experiments: DFS CS GM GS. Performed the experiments: DFS CS. Analyzed the data: DFS GS. Contributed reagents/materials/analysis tools: DFS GAC GS. Wrote the paper: DFS CS GAC GM GS.